\documentclass[pra,aps,showpacs,preprint]{revtex4}
\begin{document}  

\preprint{}

\title{Non-Gaussianity of Large-Scale Cosmic Microwave Background 
Anisotropies beyond Perturbation
Theory}

\author{Nicola Bartolo}\email{nbartolo@ictp.trieste.it}
\affiliation{The Abdus Salam International Centre for Theoretical 
Physics, Strada Costiera 11, 34100 Trieste, Italy}

\author{Sabino Matarrese}\email{sabino.matarrese@pd.infn.it}
\affiliation{Dipartimento di Fisica ``G.\ Galilei'', Universit\`{a} di Padova, 
        and INFN, Sezione di Padova, via Marzolo 8, Padova I-35131, Italy}

\author{Antonio Riotto}\email{antonio.riotto@pd.infn.it}
\affiliation{INFN, Sezione di Padova, via Marzolo 8, I-35131, Italy}

\date{\today}

\begin{abstract}
\noindent
We compute the fully non-linear Cosmic Microwave Background (CMB) 
anisotropies on scales larger than the horizon at last-scattering 
in terms of only the curvature perturbation, providing 
a generalization of the linear 
Sachs-Wolfe effect at any order in perturbation theory. 
We  show how to compute the $n$-point 
connected correlation functions of the large-scale 
CMB anisotropies for  
generic primordial seeds provided  
by standard slow-roll inflation as well as the curvaton and other scenarios 
for the generation of  cosmological perturbations.
As an application of our formalism, we compute the three- and 
four-point connected correlation functions whose detection in
future CMB experiments might be  used to assess the level of 
primordial non-Gaussianity, giving the theoretical predictions for the 
parameters of quadratic and cubic non-linearities 
$f_{\rm NL}$ and $g_{\rm NL}$.

\noindent

\end{abstract}


\maketitle

\section{Introduction}
\noindent
Cosmological inflation \cite{lrreview} is the dominant paradigm to 
understand the initial conditions for CMB anisotropies and structure 
formation. In the inflationary picture, 
the primordial cosmological perturbations are
created from quantum fluctuations ``redshifted'' out of the horizon during an
early period of superluminal expansion of the universe, where they
remain ``frozen''. They are observable as temperature 
anisotropies in the CMB at the last scattering surface. They were
first detected by the Cosmic Background 
Explorer (COBE) satellite \cite{smoot92,bennett96,gorski96}.
The last and most impressive confirmation of the inflationary paradigm has 
been recently provided by the data 
of the Wilkinson Microwave Anisotropy Probe (WMAP) 
mission~\cite{wmap1}.
Since the observed cosmological perturbations are of the order
of $10^{-5}$, one might think that first-order perturbation theory
will be adequate for all comparison with observations. That may not be
the case however, as the Planck satellite \cite{planck} and its
successors may be sensitive to the non-Gaussianity of the cosmological
perturbations at the level of second- or higher-order perturbation theory
\cite{review}. Statistics like the bispectrum and the trispectrum of the 
CMB can be used to assess the level of primordial non-Gaussianity on various 
cosmological scales and to discriminate it from the one induced by 
secondary anisotropies and systematic effects \cite{review,hu,dt,jul}. 
Therefore, it is of fundamental importance to provide accurate
theoretical predictions for the statistics of the large-angle
CMB anisotropies as left imprinted by the primordial
seeds originated during or
immediately after inflation. Steps towards this goal have been 
taken in Refs. \cite{pc,mm,prl,tomita} at the level of second order
perturbation theory.

In this paper we derive an expression for the anisotropies of the CMB  
on scales larger than the horizon at last scattering which is 
valid at any order in perturbation theory, 
providing a  fully non-linear generalization to the  
Sachs-Wolfe effect at first- \cite{sw} and second-order \cite{prl}.
In particular, for the 
standard single-field models of inflation, 
we provide the exact non-perturbative expression for the
bispectrum and the trispecturm in the so-called ``squeezed'' limit
in which some of the wavenumbers are much smaller than the others. 
Furthermore, we compute the generic expressions for the
non-linearity parameters $f_{\rm NL}$ and
$g_{\rm NL}$ characterizing respectively the quadratic and cubic
non-linearity in the large-angle CMB anisotropies.

The paper is  organized as follows. In Section II we provide the
non-linear generalization of the Sachs-Wolfe effect which is expressed in terms
of the comoving curvature perturbation from inflation in section
III. In Section IV we show how to compute 
the connected $n$-point correlation functions of the CMB anisotropies, leaving
the details to the Appendix. Finally, we draw our conclusions in Section V.
 
\section{The non-linear Sachs-Wolfe effect}
\noindent  
Our starting point is the Arnowitt-Deser-Misner
(ADM) formalism which is particularly useful to deal with the 
non-linear evolution of cosmological perturbations. The line element is 
\begin{equation}
ds^2=-N^2\,dt^2+N_i\,dt\,dx^i+\gamma_{ij}\,dx^i\,dx^j \, , \nonumber
\end{equation}
where the three-metric $\gamma_{ij}$, the lapse $N$ and the shift 
$N_i$ functions describe the evolution of  timelike hypersurfaces.
In the ADM formalism
the equations simplify considerably if we set $N^i=0$. Moreover we are  
interested only in scalar perturbations in a flat Universe and therefore we 
find it convenient 
to recast the metric as   
\begin{equation}
\label{basicmetric}
ds^2=-e^{2 \Phi} \,dt^2 + a^{2}(t) e^{-2 \Psi} \delta_{ij} \,dx^i\,dx^j\, ,
\end{equation} 
where $a(t)$ is the scale factor describing the evolution of the
homogeneous and isotropic Universe  and we have introduced  
two gravitational potentials $\Phi$ and $\Psi$. The
expression (\ref{basicmetric}) 
holds at any order in perturbation theory. To make contact
with the usual perturbative approach, one may 
expand the gravitational potentials 
 at first- and second-order, {\it e.g}, 
$\Phi=\Phi_1+\Phi_2/2$.  From Eq.~(\ref{basicmetric}) 
one recovers at linear order the well-known  
longitudinal gauge, $ N^2=(1+2\Phi_1)$ and $\gamma_{ij}
=a^2(1-2\Psi_1)
\delta_{ij}$. 
At second-order, one finds 
$\Phi_2=\phi_2-2 \phi_1^2$ and $\Psi_2=
\psi_2+2\psi_1^2$ where $\phi_1$, $\psi_1$ 
and $\phi_2$, $\psi_2$ (with 
$\phi_1=\Phi_1$ and $\psi_1=\Psi_1$)
are the first and second-order gravitational 
potentials in the longitudinal (Poisson) gauge adopted in 
  Refs.~\cite{MMB,review}, $N^2=(1+2\phi_1+\phi_2)$ and
$\gamma_{ij}=a^2(1-2\psi_1-\psi_2)
\delta_{ij}$ as far as  scalar perturbations are concerned.
In writing Eq.~(\ref{basicmetric}) we have neglected vector and tensor 
perturbation modes. For the vector perturbations the reason is that we are 
interested in long-wavelength perturbations, {\it i.e.}  
on scales larger than the horizon at last 
scattering, while vector modes will contain gradient terms being produced 
as non-linear combination of scalar-modes and thus they will be more important 
on small scales (linear vector modes 
are not generated in standard mechanisms for 
cosmological perturbations, as inflation). For example the results of 
Ref.~\cite{MMB} show clearly this for second-order perturbations. 
The tensor contribution 
can be neglected for two reasons. First, the tensor perturbations    
produced from inflation on large scales give a negligible contribution to 
the higher-order statistics of the Sachs-Wolfe effect
being of the order of (powers of) the slow-roll 
parameters during inflation (this holds for linear tensor modes as well as for 
tensor modes generated by the non-linear evolution of scalar perturbations 
during inflation, for example see the results of 
Ref.~\cite{maldacena} 
for second-order perturbations). Moreover, while on 
large scales the tensor modes have been 
proven to remain constant in time~\cite{Salopek1}, when they approach the 
horizon they have a wavelike contribution which oscillates with decreasing 
amplitude.                 

Since we are interested  in   the cosmological
perturbations on large scales, that is in 
perturbations whose wavelength is  larger than the Hubble radius at last 
scattering, a local
observer would see them in  the form of  a classical -- possibly 
time-dependent -- 
(nearly zero-momentum) homogeneous and isotropic
background. Therefore,  it should be
possible to perform a change of coordinates in such a way as to
absorb  the  
super-Hubble modes and work with a metric
of an homogeneous and isotropic Universe (plus, of course, cosmological
perturbations on  scale smaller than the horizon). 
We  split the gravitational potential $\Phi$ as
\begin{equation}
\Phi=\Phi_\ell +\Phi_s\, ,
\end{equation}
where $\Phi_\ell$ stands for the part of the
gravitational potential receiving contributions only from the
super-Hubble modes; $\Phi_s$ receives contributions only
from the sub-horizon modes

\begin{eqnarray}
\Phi_\ell&=&\int\frac{d^3\!k}{(2\pi)^3}\, \theta\left(aH-k\right)
\, \Phi_{\vec{k}} \ e^{i\vec{k}\cdot\vec{x}} \, ,\nonumber\\
\Phi_s&=&\int\frac{d^3\!k}{(2\pi)^3}\, \theta\left(k-aH\right)
\, \Phi_{\vec{k}} \ e^{i\vec{k}\cdot\vec{x}} \, ,
\end{eqnarray}
where $H$ is the Hubble rate computed with respect to the cosmic time,
$H=\dot{a}/a$, and $\theta(x)$ is the step function. Analogous definitions
hold for the other gravitational potential $\Psi$. 

By construction $\Phi_\ell$ and $\Psi_\ell$  are a collection of 
Fourier modes whose wavelengths are larger than the horizon
length and we may safely neglect  their spatial gradients.
Therefore $\Phi_\ell$ and 
$\Psi_\ell$ are only  functions of time.
This amounts to saying that  
we can absorb the large-scale perturbations in the metric (\ref{basicmetric})
by the following redefinitions
\begin{eqnarray}
\label{tbar}
d\overline{t}&=&e^{\Phi_\ell} dt\, , \\
\label{abar}
\overline{a}&=&a\, e^{-\Psi_\ell}\, .
\end{eqnarray}
The new metric  describes a  homogeneous and 
isotropic Universe
\begin{equation}
\label{newmetric}
ds^2=-d\overline{t}^2 + \overline{a}^{2} \delta_{ij} \,dx^i\,dx^j\, ,
\end{equation} 
where for simplicity we have not included  the sub-horizon modes.
On super-horizon scales one can regard  the Universe as a collection
of regions of size of the Hubble radius 
evolving like  unperturbed patches with metric (\ref{newmetric})
~{\cite{Salopek1}.

Let us now go back to  the quantity we are interested in, namely the 
anisotropies of the CMB as measured today by an observer ${\mathcal O}$.
If she/he is interested in the CMB anisotropies at large scales, the effect
of super-Hubble modes is encoded in the metric (\ref{newmetric}). 
During their travel    
from the last scattering surface -- to be considered as the  
emitter point ${\mathcal E}$ --  to the observer,  
the CMB photons suffer a redshift 
determined by the ratio of the emitted 
frequency $\overline{\omega}_{\mathcal E}$ to the observed one 
$\overline{\omega}_{\mathcal O}$ 
\begin{equation}
\label{Texact}
\overline{T}_{\mathcal O}=\overline{T}_{\mathcal E}\,
\frac{\overline{\omega}_{\mathcal O}}{\overline{\omega}_{\mathcal E}}\, , 
\end{equation}
where $\overline{T}_{\mathcal O}$ and 
$\overline{T}_{\mathcal E}$ are  the temperatures at the
observer point and  at the
last scattering surface, respectively.

What is then the temperature anisotropy measured by the  observer?
The expression (\ref{Texact}) shows that 
the measured large-scale anisotropies  
are made of two contributions:  the intrinsic 
inhomogeneities in the temperature at the last scattering
surface  and the inhomogeneities in the scaling factor
provided by the ratio of the frequencies of the photons
at the departure and arrival points. 
Let us first consider the second contribution.
As the 
frequency of the photon  is the inverse of a time period, we get
immediately the fully non-linear relation
\begin{equation}
\label{1result}
\frac{\overline{\omega}_{\mathcal E}}{\overline{\omega}_{\mathcal O}}
=\frac{\omega_{\mathcal E}}{\omega_{\mathcal O}}e^{-\Phi_{\ell
{\mathcal E}}+\Phi_{\ell
{\mathcal O}}}\, .
\end{equation}             
As for the   temperature anisotropies coming from the intrinsic 
temperature fluctuation at the emission point, 
it maybe worth to recall   how to obtain this quantity 
in the longitudinal gauge at first order. By expanding the photon energy
density $\rho_\gamma \propto T_\gamma^4$, the intrinsic temperature 
anisotropies at last scattering are given by  $\delta_1 T_{\mathcal E}/
T_{\mathcal E}=(1/4)\delta_1 
\rho_\gamma/\rho_\gamma$. One relates the photon energy density 
fluctuation to the gravitational perturbation first by implementing the 
adiabaticity condition $\delta_1 
\rho_\gamma/\rho_\gamma=(4/3)\delta_1 
\rho_m/\rho_m$, 
where $\delta_1\rho_m/\rho_m$ 
is the relative fluctuation in the matter component, 
and then using the energy constraint of Einstein equations 
$\Phi_1=-(1/2)\delta_1\rho_m/\rho_m$. The result is $\delta_1 T_{\mathcal E}/
T_{\mathcal E}=-2 \Phi_{1\mathcal E}/3$. Summing this contribution
to the anisotropies coming from 
the redshift factor (\ref{1result}) expanded at first order provides the  
standard (linear) Sachs-Wolfe effect 
$\delta_1 T_{\mathcal O}/T_{\mathcal O}
=\Phi_{1\mathcal E}/3$. Following the same steps,  
we may easily obtain its full non-linear generalization.   

Let us first relate the photon energy density 
$\overline{\rho}_\gamma$ to the energy density of the non-relativistic matter 
$\overline{\rho}_m$ by using the adiabaticity conditon. 
Again here  a bar indicates  that we are considering  
quantities in the locally homogeneous Universe described by the metric
(\ref{newmetric}). Using the 
energy continuity equation on large scales $\partial \overline{\rho}/  
\partial\overline{t} =-3 \overline{H} (\overline{\rho}+\overline{P})$, where 
$\overline{H}=d \ln \overline{a}/d \overline{t}$ 
and $\overline{P}$ is the pressure of the fluid, one can easily show that 
there exists a conserved 
quantity in time at any order in perturbation theory ~\cite{KMNR}
\begin{equation}
{\cal F} \equiv \ln \overline{a} +\frac{1}{3}\int^{\overline{\rho}}\, 
\frac{d \overline{\rho'}}{\left(\overline{\rho'}+\overline{P'}\right)}\, .
\label{fun}
\end{equation}
The perturbation $\delta {\cal F}$ is a gauge-invariant quantity representing 
the non-linear extension of the curvature perturbation $\zeta$ 
on uniform energy density hypersurfaces on superhorizon scales 
for adiabatic fluids~\cite{KMNR}. Indeed, expanding it at 
first and second order one gets the corresponding definition 
$\zeta_1=-\psi_1-\delta_1\rho/ \dot{\rho}$ and the quantity 
$\zeta_2$ introduced in Ref.~\cite{MW}. 
At first order the adiabaticity condition 
corresponds to set $\zeta_{1\gamma}=\zeta_{1 m}$ for the curvature 
perturbations relative to each component. At the non-linear level
the adiabaticity condition generalizes to 

\begin{equation}
\frac{1}{3} \int \frac{d \overline{\rho}_m}{\overline{\rho}_m} =
\frac{1}{4} \int  
\frac{d \overline{\rho}_\gamma}{\overline{\rho}_\gamma}\, ,
\end{equation}
or
\begin{equation}
\ln \overline{\rho}_m = \ln \overline{\rho}_\gamma^{3/4}\, .
\label{2result}
\end{equation}
To make contact with the standard second-order result, we may 
expand  in Eq.~(\ref{2result}) the photon energy density 
perturbations as 
$\delta \overline{\rho}_\gamma/\rho_\gamma=
\delta_1 {\rho}_\gamma/ \rho_\gamma+\frac{1}{2} 
\delta_2 {\rho}_\gamma/\rho_\gamma $, and similarly for the matter 
component. We   immediately recover 
the adiabaticity condition  

\begin{equation}
\frac{\delta_2 {\rho}_\gamma}{\rho_\gamma}=
\frac{4}{3} \frac{\delta_2 {\rho}_m}{\rho_m}+
\frac{4}{9} \left(\frac{\delta_1 {\rho}_m}{\rho_m}\right)^2
\end{equation}
given in
Ref. \cite{review}.
 
Next we need to relate the photon energy density to the gravitational 
potentials at the non-linear level. The energy 
constraint inferred from the (0-0) component of Einstein equations in the 
matter-dominated era with the  
 ``barred'' metric~(\ref{newmetric}) is
\begin{equation}
\label{0-0}
\overline{H}^2=\frac{8\pi G_N}{3} \overline{\rho}_m\, .
\end{equation}     
Using Eqs.~(\ref{tbar}) and~(\ref{abar}) the 
Hubble parameter $\overline{H}$ reads
\begin{equation}
\overline{H}=\frac{1}{\overline{a}}\frac{d\overline{a}}{d \overline{t}}=
e^{-\Phi_\ell}
(H-\dot{\Psi}_\ell)\, ,
\end{equation} 
where $H=d \ln a/dt$ is the  Hubble parameter in the
``unbarred'' metric.   
Eq.~(\ref{0-0}) thus yields an expression for the energy density of the 
non-relativistic matter which is fully nonlinear, being expressed in terms of 
the gravitational potential $\Phi_\ell$
\begin{equation}
\label{3result}
\overline{\rho}_m=\rho_m e^{-2 \Phi_\ell}\, ,
\end{equation}
where we have dropped $\dot{\Psi}_\ell$ which is negligible on 
large scales. 
By perturbing the expression ~(\ref{3result}) we are able to 
recover in a straightforward way the solutions
of  the (0-0) component of Einstein equations for 
a matter-dominated 
Universe in the large-scale limit obtained at 
second-order in perturbation theory. Indeed,  recalling that 
$\Phi$ is perturbatively related 
to the quantity $\phi=\phi_1+\phi_2/2$ used in Ref.~\cite{review} by 
$\Phi_1=\phi_1$ and $\Phi_2=\phi_2-2 (\phi_1)^2$,  one immediately  
obtains  \cite{review,BMR4}

\begin{eqnarray}
\frac{\delta_1\rho_m}{\rho_m} &=& -2 \phi_1\, ,\nonumber\\  
\frac{1}{2}\frac{\delta_2 \rho_m}{\rho_m}&=&- \phi_2+4 (\phi_1) ^2\, .
\end{eqnarray}
The expression for the intrinsic temperature of the photons at 
the last scattering surface $\overline{T}_{\mathcal E} 
\propto \overline{\rho}^{1/4}_\gamma$ follows from  
Eqs.~({\ref{2result}) and (\ref{3result})  
\begin{equation}
\label{4result}
\overline{T}_{\mathcal E}=T_{\mathcal E}\, e^{-2 \Phi_\ell/3}\, .
\end{equation}
Plugging Eqs.~(\ref{1result}) and (\ref{4result}) into 
the expression (\ref{Texact}) we are finally able 
to provide the expression for the 
CMB temperature which is fully nonlinear and takes 
into account both the gravitational redshift of the photons due to the 
metric perturbations at last scattering and the intrinsic temperature 
anisotropies 
\begin{equation}
\label{final}
\overline{T}_{\mathcal O}=\left(
\frac{\omega_{\mathcal O}}{\omega_{\mathcal E}}\right) 
T_{\mathcal E}\, e^{\Phi_\ell/3}\, .
\end{equation}
From Eq.~(\ref{final}) we read the {\it non-perturbative} 
anisotropy corresponding to the Sachs-Wolfe effect
\begin{equation}
\label{final2}
\frac{\delta_{np} \overline{T}_{\mathcal O}}{T_{\mathcal O}}=
e^{\Phi_\ell/3}-1\, .
\end{equation}        
Eq.~(\ref{final2}) is  one of the main results of this paper and 
represents
{\it at any order in perturbation theory} the extension of the 
linear Sachs-Wolfe effect. At first order one gets
\begin{equation}
\frac{\delta_1 T_{\mathcal O}}{T_{\mathcal O}}=\frac{1}{3} \Phi_1\, ,
\end{equation} 
and at  second order 
\begin{equation}
\label{SW2nd}
\frac{1}{2} \frac{\delta_2 T_{\mathcal O}}{T_{\mathcal O}}=
\frac{1}{6} \Phi_2+\frac{1}{18} \left( \Phi_1 \right)^2\, ,
\end{equation}
which exactly reproduces 
the generalization of the Sachs-Wolfe effect at second-order in the 
perturbations found in Ref.~\cite{review,BMR4} (where 
$\Phi_1=\phi_1$ and $\Phi_2=\phi_2-2 (\phi_1)^2$). 

\section{Relating the CMB anisotropies to the inflationary comoving
curvature perturbation}
\noindent
In this section we relate the gravitational potentials 
$\Phi_\ell$ to $\Psi_\ell$ to the curvature 
perturbation $\zeta=\delta {\mathcal F}$ at any order in perturbation theory
(for notational simplicity we drop the subscrip ``$\ell$'' from now on).
This will allow to express the non-linear temperature fluctuations
in terms of the initial conditions provided by inflation. 
We use the evolution equation in the ADM formalism
\begin{equation}
\label{i-jADM}
-N^{|i}_{\;\; |k}+\frac{1}{3} N^{|l}_{\;\; |l} \delta^{i}_{\; k}+N\,\,
^{(3)}\overline{R}^{i}_{\; k}=N\,\, 8\pi G\,\, \overline{S}^{i}_{\;k}\, , 
\end{equation}
where a vertical bar denotes a covariant derivative 
with respect to $\gamma_{ij}$ and 
we have used that fact that the traceless part of the extrinsic curvature 
$\overline{K}_{ij}=K_{ij}-K\,\gamma_{ij}/3$ vanishes in 
our metric~(\ref{newmetric}). The extrinsic curvature is defined as 
$K_{ij}=-\dot{\gamma_{ij}}/(2N)$ and $K=\gamma^{ij}K_{ij}$. Here 
$^{(3)}\overline{R}^{i}_{\;k}=\gamma^{ij}~^{(3)}\overline{R}_{jk}=
^{(3)}R^{i}_{\;k}-^{(3)}R\,\,\delta^{i}_{\; k}/3$ where
$^{(3)}R_{jk}$ 
is the Ricci tensor of constant time hypersurfaces associated with the metric 
$\gamma_{ij}$. Analogous definitions hold for $\overline{S}^{i}_{\;k}$ 
constructed from the matter stress-energy three-tensor $S_{ij}=T_{ij}$. 

From 
Eq.~(\ref{i-jADM}) we want to obtain a constraint between the gravitational 
potentials $\Phi$ and $\Psi$ keeping 
track of non-local terms. Thus we cannot neglect the gradient terms 
appearing in this equation. Since $N=e^{\Phi}$ we find
\begin{eqnarray}
N^{|i}_{\;\; |k}&=&\frac{e^{2\Psi} e^{\Phi}}{a^{2}(t)} \Big(\Phi_{,k}\Phi^{,i}+
\Phi^{,i}_{\;\;,k}+\Psi_{,k}\Phi^{,i}+\Psi^{,i}\Phi_{,k} 
-\Psi_{,j}\Phi^{,j} \delta^{i}_{\;\;k} \Big) \, ,\nonumber \\
^{(3)}R^{i}_{\;\; k}&=&
\frac{e^{2\Psi}}{a^{2}(t)}\Big( \Psi^{,i}_{\;\;,k}+\nabla^2\Psi 
\delta^{i}_{\;\;k}+\Psi^{,i}\Psi_{,k}
-\Psi_{,m}\Psi^{,m}\delta^{i}_{\;\; k}\Big)\, ,\nonumber \\
^{(3)}R&=&2\frac{e^{2\Psi}}{a^{2}(t)} 
\left(2\nabla^2\Psi-\Psi_{,i}\Psi^{,i}\right)\, .
\end{eqnarray}
Hence Eq.~(\ref{i-jADM}) reads
\begin{eqnarray}
\label{incompl}
&-&\Big(\Phi_{,k}\Phi^{,i}+
\Phi^{,i}_{\;\;,k}+\Psi_{,k}\Phi^{,i}+\Psi^{,i}\Phi_{,k} 
-\Psi_{,j}\Phi^{,j} \delta^{i}_{\;\;k} \Big)
+\frac{1}{3}\big(\nabla^{2} \Phi
+ \Phi^{,l}\Phi_{,l}-\Psi_{,l}\Phi^{,l}\Big) \delta^{i}_{\;\;k}\nonumber \\
&+&\Big(\Psi^{,i}_{\;\;,k}+\nabla^2\Psi 
\delta^{i}_{\;\;k}+\Psi^{,i}\Psi_{,k}
-\Psi_{,m}\Psi^{,m}\delta^{i}_{\;\; k}\Big)
-\frac{2}{3}\Big(2\nabla ^2 \Psi-
\Psi_{,l}\Psi^{,l} \Big)\delta^{i}_{\;\;k}=8\pi G\, a^{2}(t) e^{-2 \Psi}
\overline{S}^i_{\;k}\, , \nonumber \\
\end{eqnarray}
where the indices of the partial derivatives are raised by $\delta^{ij}$. 
Notice that at first order this equation gives the usual constraint 
$\Phi_1=\Psi_1$. As far as the matter content is concerned we can 
consider the 
perfect fluid energy momentum tensor  $T_{\mu\nu}=
\left(\overline{\rho}+\overline{P}\right)
\,u_\mu u_\nu +\overline{P}\,g_{\mu\nu}$ and we find  

\begin{eqnarray}
S^i_{\;k}&=&\gamma^{ij}T_{jk}=a^{-2}(t) e^{2\Psi} \left(\overline{\rho}
+\overline{P}\right){u^i u_k} +
\overline{P}\delta^i_{\;k}\, ,\nonumber\\
\overline{S}^i_{\;k}&=&a^{-2}(t)e^{2\Psi}
\left(\overline{\rho}+\overline{P}\right)(u^iu_k-
u^iu_i\delta^{i}_{\;k}/3)\, ,
\end{eqnarray}
where
$u^i=\delta^{ij}u_j$. We need an expression for the spatial velocities. Thus 
we use the momentum constraint of the ADM equations which reads~\cite{Bardeen1}
\begin{equation}
\label{momentumconstr}
8 \pi G\, J_i=-\frac{2}{3} K_{|i}\, ,
\end{equation}
where $J_{i}=NT^{0}_{\;\;i}=e^{\Phi}T^{0}_{\;\;i}=-e^{-\Phi}T_{0i}$ 
is the momentum density and $K=-3e^{-\Phi} (H(t)-\dot{\Psi})$. 
One can use the normalization 
$g^{\mu\nu}u_\mu u_\nu=-1$ to express $u_0$, finding
$u_{0}=-(1+a^{-2}(t) e^{2\Psi} u^i u_i)^{1/2}$. Let us just consider scalar 
velocities $u^i=\partial_i u$ (on sufficiently large scales
vector modes may be safely neglected). 
From Eq.~(\ref{momentumconstr}) we obtain
\begin{equation}
\label{momentumconstr2}
(\overline{\rho}+\overline{P}) 
u_i(1+a^{-2}(t) e^{2\Psi} u^i u_i)^{1/2}
=\Big[\frac{e^{-\Phi}}{4 \pi G}(H(t)-\dot{\Psi}) \Big]_{|i}
\end{equation}
In fact we are interested in the case of non-relativistic matter 
$\overline{P}_m=0$ and $\overline{\rho}_m=
\rho_m e^{-2\Phi}$.
As  
Eq.~(\ref{incompl}) contains at least two gradients (also 
$\overline{S}^i_k$ contains at least two spatial gradients), using the
 gradient 
expansion we may restrict ourselves to  the   solution
\begin{equation}
u_i=\frac{e^{2 \Phi}}{4\pi G\, \rho_m} 
\Big[e^{-\Phi}(H(t)-\dot{\Psi}) \Big]_{|i}=-\frac{e^{\Phi} 
H(t)}{4\pi G\,\rho_m}
\Phi_{,i}\, ,
\end{equation}
where we have neglected $\dot{\Psi}$. We thus find
\begin{equation}
\label{Sik}
\overline{S}^i_{\;k}=\frac{e^{2\Psi}}
{6 \pi G\, a^{2}} \Big(\Phi^{,i}\Phi_{,k}-\frac{1}{3}
\Phi^{,j}\Phi_{,j} \delta^i_{\;k} \Big)\, .
\end{equation}
Inserting Eq.~(\ref{Sik}) into Eq~(\ref{incompl}) and applying the operator 
$\partial_i \partial^k$ we find 
\begin{eqnarray}
\label{constr}
\nabla^4 \left(\Psi-\Phi\right)&=&-\frac{3}{2}\partial_i\partial^k
\left( \Psi^{,i}\Psi_{,k} \right)+\frac{1}{2}\nabla^2\left(\Psi^{,i}\Psi_{,i} 
\right) \nonumber \\
&+&\frac{7}{2}\partial_i\partial^k
\left( \Phi^{,i}\Phi_{,k} \right) 
-\frac{7}{6}\nabla^2
\left(\Phi^{,i}\Phi_{,i} \right) \nonumber \\
&+&3 \partial_i\partial^k
\left( \Phi^{,i}\Psi_{,k} \right)-\nabla^2 \left(\Phi^{,i}\Psi_{,i} 
\right) \, .
\end{eqnarray}
Notice that Eq.~(\ref{constr}) is the non-linear generalization of the  
constraint between the gravitational potentials $\Phi$ and $\Psi$ in the 
longitudinal (Poisson) gauge and is valid at any order
in perturbation theory. At linear 
order one recovers the 
well-known result $\Phi_1=\Psi_1$, while at second order the relation 
first found in 
Refs.~\cite{enhanc,review} follows $\psi_2-\phi_2=-4 \psi_1^2 +10 \nabla^{-4} 
\partial_i\partial^k  (\psi_1^{,i} \psi_{1,k} )-\frac{10}{3} \nabla^{-2}
(\psi_1^{,i}\psi_{1,i})$ (where $\Phi_2=\phi_2-2 \phi_1^2$ and $\Psi_2=
\psi_2+2\psi_1^2$ and  $\phi_1=\Phi_1$, $\psi_1=\Psi_1$). 

In the following we find it convenient to write the relation~(\ref{constr}) as 
$\Psi=\Phi+{\mathcal K}[\Phi,\Psi]$ where the kernel ${\mathcal K}$ 
is obtained by acting through the  operator  $\nabla^{-4}$ onto 
the {\it r.h.s.} of Eq.~(\ref{constr}) and 
takes into account non-local terms. Going to momentum space, one 
easily  realizes that in 
the specific ``squeezed'' limit of 
Ref.~\cite{maldacena}, where one of the 
wavenumbers is much smaller than the other two, 
{\it e.g.} $k_1 \ll k_{2,3}$, the kernel
${\mathcal K} \rightarrow 0$. From Eq.~(\ref{fun}) 
the curvature perturbation is given by
\begin{equation}
\label{zetaphipsi}
\zeta\equiv\delta {\mathcal F}=-\Psi+\frac{1}{3} 
\ln \frac{\overline{\rho}}{\rho}=-\Psi-\frac{2}{3} \Phi \, ,
\end{equation} 
where we have used $\overline{\rho}_m=\rho_m e^{-2 \Phi}$. Hence we finally 
find $-5 \Phi/3=\zeta+{\mathcal K}$ and we can recast the non-linear 
temperature anisotropies~(\ref{final2}) as 
\begin{equation}
\label{final3}
\frac{\delta_{np} \overline{T}_{\mathcal O}}{T_{\mathcal O}}=
e^{-\zeta/5-{\mathcal K}/5}-1\, ,
\end{equation} 
which is the starting point to evaluate 
the $n-$point correlation function for  CMB temperature anisotropies.
The comoving curvature perturbation $\zeta$ is conserved at any order in
perturbation theory on large scales and therefore one can fix its properties
right at the end of inflation. Since for 
standard inflation the curvature perturbation can be considered as a  
Gaussian distributed quantity \cite{acquaviva,maldacena}
(deviations from non-Gaussianity are proportional to deviations
of the spectral index from unity and are, therefore, tiny), we will
adopt $\zeta$ as our Gaussian seed in the case of standard
single-field models of inflation.

\section{Correlation functions of large-scale CMB anisotropies}
\noindent
As a  warm-up excercise, let us first consider a  simpler case and
 evaluate the $n$-point 
correlation function of $e^{\varphi({\bf x})}$ where $\varphi({\bf x})$ is a 
Gaussian random field. Applying the well-known  techniques of quantum 
field theory, it turns out that 
\begin{equation}
\label{path}
\langle e^{\varphi({\bf x}_1)} \,.....\,e^{\varphi({\bf x}_N)} \rangle = 
e^{\frac{1}{2} \int d{\bf x}d{\bf y} J({\bf x})\, 
\langle \varphi({\bf x})\varphi({\bf y}) \rangle\, J({\bf y})}\, ,
\end{equation}
where $J({\bf x})=\sum_{i=1}^N \delta({\bf x}-{\bf x}_i)$ corresponds in fact 
to the source term appearing in the path integral formulation. 
The calculation of the integral in Eq.~({\ref{path}) brings
\begin{equation}
\label{casoK=0}
\langle e^{\varphi({\bf x}_1)} \,.....\,e^{\varphi({\bf x}_N)} \rangle =  
e^{\frac{1}{2} \sum_{i, j} \langle  \varphi({\bf x}_i)  
\varphi({\bf x}_j) \rangle}\, .
\end{equation}
Notice, for example, that for the 2-point correlation function we find the 
usual result
\begin{equation}
\langle e^{\varphi({\bf x}_1)} e^{\varphi({\bf x}_2)} \rangle = 
e^{\langle \varphi^2 \rangle} e^{\langle \varphi({\bf x}_1) 
\varphi({\bf x}_2)\rangle } \, ,
\end{equation}
which for ${\bf x}_1={\bf x}_2$ gives 
$\langle e^{2 \varphi({\bf x})} \rangle=e^{2 \langle \varphi^2\rangle }$,
where $\langle \varphi^2 \rangle=\langle \varphi^2({\bf x}) \rangle$.

For the 3-point function one finds
\begin{equation}
\langle e^{\varphi({\bf x}_1)} e^{\varphi({\bf x}_2)} 
e^{\varphi({\bf x}_3)}\rangle = e^{\frac{3}{2} \langle \varphi^2 \rangle}\, 
e^{\langle \varphi_1 \varphi_2 \rangle + \langle 
\varphi_1 \varphi_3 \rangle + \langle \varphi_2 \varphi_3 \rangle }\, , 
\end{equation}
where for simplicity we have used 
the notation $\langle \varphi_i \varphi_j \rangle=
\langle \varphi({\bf x}_i) \varphi({\bf x}_j) \rangle $. If now 
we expand the exponential in the limit in which the two point 
function is small
we obtain
\begin{eqnarray}
\label{giusta?}
\langle e^{\varphi({\bf x}_1)} e^{\varphi({\bf x}_2)} 
e^{\varphi({\bf x}_3)}\rangle &\simeq& 
e^{\frac{3}{2} \langle \varphi^2 \rangle}\Big[1
+\Big( \langle \varphi_1 \varphi_2 \rangle+{\rm cycl.}\Big) \nonumber \\
&+&
\frac{1}{2}\left( \langle \varphi_1 \varphi_2 \rangle^2+{\rm cycl.} \right)
\nonumber \\
&+&\Big( \langle \varphi_1\varphi_2 \rangle \langle \varphi_1\varphi_3 \rangle
+{\rm cycl.} \Big) \Big]\, .
\end{eqnarray}
It is in fact a term analogous to 
the last contribution in Eq.~(\ref{giusta?}) 
that enters in the three-point function of the CMB 
anisotropies~(\ref{final2}) if we are 
interested in non-linearities up to second-order terms only. The
complication arises from the fact that the 
gravitational potential $\Phi$ appearing in Eq.~(\ref{final2})  is not 
a Gaussian variable, since it can already contain quadratic 
(and higher order) terms in the Gaussian variable $\zeta$ in the case of single-field models of inflation. 
These will add  non-Gaussian contributions which are  contained in the
 kernel ${\mathcal K}$. In other scenarios for the generation of the 
cosmological perturbations $\zeta$ is not a Gaussian quantity and 
we will properly take into account 
the primordial non-Gaussian contributions. 

Let us see how to generalize the previous 
procedure to this case. First of all we apply an iterative procedure to 
express the kernel ${\mathcal K}$ in terms of powers of ${\zeta}$. 
We will use Eqs. ~(\ref{constr}) and ~(\ref{zetaphipsi}). 
For the zeroth and first-order terms of the iteration 
we find (the suffix does not refer to the order of the expansion in 
the perturbations, 
but to the order of the approximation given by the iteration procedure: each 
$r$-th term contains up to $(r+1)$ powers of $\zeta$) 
\begin{eqnarray} 
\label{0}
\Phi^{(0)}&=&-\frac{3}{5} \zeta\, \\
\label{1}
\Phi^{(1)}&=&-\frac{3}{5} \zeta -\left(\frac{3}{5}\right)^3 
\mathcal{K}[\zeta^2]
\, ,  
\end{eqnarray}
and for the next terms ($n=1,2,..$) 
\begin{eqnarray}
\Phi^{(2n)}&=&
\Phi^{(2n-1)}+{\mathcal K}_1[\Phi^{(0)},\Phi^{(2n-2)}-\Phi^{(2n-1)}]
\\
&+&\sum_{m=0}^{n-2}{\mathcal K}_1[\Phi^{(m)}-\Phi^{(m+1)}, \Phi^{(2n-m-3)}
-\Phi^{(2n-m-2)}] \nonumber\, , \\
\Phi^{(2n+1)}&=&\Phi^{(2n)}
+{\mathcal K}_1[\Phi^{(0)},\Phi^{(2n-1)}-\Phi^{(2n)}]
\nonumber \\
&+&\sum_{m=0}^{n-2}{\mathcal K}_1[\Phi^{(m)}-\Phi^{(m+1)}, \Phi^{(2n-m-2)}
-\Phi^{(2n-m-1)}]\nonumber \\
&+&{\mathcal K}_2 [(\Phi^{(n-1)}-\Phi^{(n)})^2]\, ,
\end{eqnarray}
where we have introduced the bilinear operators 

\begin{eqnarray}
\mathcal{K}_1[(\cdot),(\cdot)]& \equiv &\nabla^{-4}(
6\partial_i \partial^k [(\cdot)^{,i}(\cdot)_{,k}]-2\nabla^2[
(\cdot)^{,i}(\cdot)_{,i}])\, ,\nonumber\\
\mathcal{K}_2[(\cdot),(\cdot)]& \equiv &
-\nabla^{-4}\left(\frac{1}{2} \partial_i \partial^k \left[
(\cdot)^{,i}(\cdot)_{,k}\right]-
\frac{1}{2}\nabla^2[(\cdot)^{,i}(\cdot)_{,i}]\right)\, .
\end{eqnarray} 
Notice that for equal entries 
${\mathcal K}_1=6 {\mathcal K}/5$. If the upper limit of the sums 
appearing in these expressions turns out to be negative the sum must be 
taken to be vanishing.
  
Thus we can use 
Eq.~(\ref{0}) and~(\ref{1}) to find the next order approximations for the 
expression of $\Phi$ in terms of the Gaussian curvature perturbation $\zeta$.
For example for $\Phi^{(2)}$ up to  ${\cal O}(\zeta^3)$ contributions we have
\begin{equation}
\Phi^{(2)}=
-\frac{3}{5} \zeta -\left(\frac{3}{5}\right)^3 
\mathcal{K}[\zeta^2]+{\mathcal K}_1\left[-\frac{3}{5} \zeta,
\left(\frac{3}{5}\right)^3 \mathcal{K}[\zeta^2]\right]\, .
\end{equation}
Using this iterative procedure we express the kernel ${\mathcal K}$ in 
Eq.~(\ref{final3}) as a function of $\zeta$ and we can determine the $n$-point 
connected correlation functions for the temperature 
anisotropies~(\ref{final3}) 
accounting for the information contained in ${\mathcal K}[\zeta]$. 
In the following we outline the procedure and we give the results, while more 
details on the computation can be found in the Appendix. 

In order to obtain the $n$-point (connected) correlation functions we will 
borrow again some techniques of functional-integral analysis
from quantum field theory~\cite{Ramond, Zinn, Mosel}
(for different applications of the path-integral approach in cosmology, 
see Refs.~\cite{PW, GW, MLB, Bertschinger, MVJ,MHV}).
In particular we need the so called {\it generating functional of the 
correlation functions} $W$, which in our case can be written as
\begin{equation}
Z[J]=\int{\mathcal D}[\zeta] {\mathcal P}[\zeta] e^{i\int d{\bf x} 
J({\bf x})\, (e^{-\zeta/5-{\mathcal K}(\zeta)/5}-1)}\, .
\end{equation}
Here $J({\bf x})$ is an external source 
perturbing the underlying statistics, 
${\mathcal P}[\zeta]$ is the probability density functional which in our case 
is a Gaussian one (see Eq.~(\ref{defPG}) of the Appendix), 
and  ${\mathcal D}[\zeta]$ is a suitable measure such that 
the total probability is normalized to unity, $\int{\mathcal D}[\zeta] 
{\mathcal P}[\zeta] =1$. The kernel ${\mathcal K}[\zeta]$ will be 
given by the iterative procedure. From the generating functional one can 
obtain the correlation functions by taking the functional derivatives of 
$Z[J]$ with respect to the source $J$ evaluated at $J=0$~\cite{Zinn}
\begin{eqnarray}
\label{def11}
& & \langle (e^{-\zeta_1/5-{\mathcal K}(\zeta_1)/5}-1)...
(e^{-\zeta_n/5-
{\mathcal K}(\zeta_n)/5}-1) \rangle = \nonumber \\ 
& &i^{-n} \frac{\delta^{n} Z[J]}{\delta J({\bf x}_1)...\delta J({\bf x}_n)} 
\Bigg|_{(J=0)} \equiv  Z^{(n)}({\bf x}_1,...,{\bf x}_n) 
\, ,
\end{eqnarray}      
where  
$\zeta_i \equiv \zeta({\bf x}_i)$.

By defining the new functional $W[J]=\ln Z[J]$ one can obtain the connected 
correlation functions by the same operations 
\begin{eqnarray}
\label{def22}
& & \langle (e^{-\zeta_1/5-{\mathcal K}(\zeta_1)/5}-1)...
(e^{-\zeta_n/5-
{\mathcal K}(\zeta_n)/5}-1) \rangle_{\rm conn.}= \nonumber \\ 
& &i^{-n} \frac{\delta^{n} W[J]}{\delta J({\bf x}_1)...\delta J({\bf x}_n)} 
\Bigg|_{(J=0)} \equiv  W^{(n)}({\bf x}_1,...,{\bf x}_n) 
\, .
\end{eqnarray}   
It is clear that the complete knowledge of the statistical properties of the 
perturbations, {\it i.e.} the complete knowledge of the correlation 
functions at all orders, can be achieved if one knows the generating 
functionals. In turn from the definitions~(\ref{def11}) 
and~(\ref{def22}) the correlation functions will appear in a power series 
expansion of the  generating functionals, as 
\begin{eqnarray}
Z[J]&=&1+\sum_{n=1}^{\infty} \frac{i^n}{n!} \int d{\bf x}_1... d{\bf x}_n 
Z^{(n)}({\bf x}_1,...,{\bf x}_n)\, J({\bf x}_1)...J({\bf x}_n)\, ,\nonumber\\
W[J]&=&\sum_{n=1}^{\infty} \frac{i^n}{n!} \int d{\bf x}_1... d{\bf x}_n 
W^{(n)}({\bf x}_1,...,{\bf x}_n)\, J({\bf x}_1)...J({\bf x}_n)\, .
\end{eqnarray}

To compute the functional derivatives of $W[J]$ one can follow the 
standard procedure to evaluate the connected correlation functions used in 
field theory \cite{Ramond}. The computation makes use of a 
perturbative expansion around some known solution, 
which corresponds to the connected 
correlation functions of the free scalar field. In our case, 
the known solution corresponds to the 
correlation functions when the kernel ${\mathcal K}$ vanishes, which we have 
computed previously in Eq.~(\ref{casoK=0}). As  explained in detail 
in  the 
Appendix,  we perform the expansion 
around ${\mathcal K}=0$ using as
 expansion parameter the {\it r.m.s}  amplitude of 
the cosmological perturbations themselves, that is  
$\langle \zeta^2 \rangle^{1/2} \sim 10^{-5}$. 

\subsection{The Bispectrum}
\noindent
From the general results presented in  the Appendix we provide now
the expression for the 3-point connected correlation function. It suffices 
to expand 
the kernel ${\mathcal K}$ up to second order.  
The total
kernel can be written  
as a convolution in configuration space
\begin{equation}
\label{Kexpansion3p}
{\mathcal K}(\zeta)=\int d{\bf x}_1 d{\bf x}_2 K_2({\bf x}
-{\bf x}_1, {\bf x}-{\bf x}_2) \zeta({\bf x}_1)  
\zeta({\bf x}_2)\, ,
\end{equation}
where $
K({\bf x}-{\bf x}_1, {\bf x}-{\bf x}_2)$ is the double
inverse Fourier transform of the expression 

\begin{equation}
\label{Kexpansion3p1}
\widetilde{K}_2\left({\bf k}_1,{\bf k}_2\right)=\left(a_{\rm NL}-1\right)
+
\frac{9}{5}\left[({\bf k}_1\cdot{\bf k}_3)
({\bf k}_2\cdot{\bf k}_3)/k^4-(1/3)({\bf k}_1\cdot{\bf k}_2)/k^2\right]\, ,
\end{equation}
where 
$k=\left|{\bf k}_3\right|$ and ${\bf k}_3=-({\bf k}_1+{\bf k}_2)$.
In Eq. (\ref{Kexpansion3p}) we have added 
the constant  $a_{\rm NL}$ whose role is to 
parametrize the primordial non-Gaussianity
generated during or after inflation in the various scenarios for the
generation of the cosmological perturbations, 
$\zeta=\zeta_{\rm L}+ (a_{\rm NL}-1) 
\star \zeta_{\rm L}^2$ (from now on we will remove the subscript
``L''). For instance,
in single field models of inflation $a_{\rm NL}=1$ (plus tiny contributions
proportional to the deviation from scale invariance), in the
curvaton   scenarion $a_{\rm NL}=(3/4r)-r/2$,  where $r\simeq
(\rho_\sigma/\rho)_D$ represents the relative curvaton contribution to the
total energy density at curvaton decay \cite{review}. 

We find
\begin{eqnarray}
\label{3point}
W^{(3)}({\bf x}_1, {\bf x}_2, {\bf x}_3)&=& 
\langle (e^{-\zeta({\bf x}_1)/5}-1) 
(e^{-\zeta({\bf x}_2)/5}-1) (e^{-\zeta({\bf x}_3)/5}-1) 
\rangle_{\rm connected}
\nonumber \\
&-& 5 \sum_p   
\int d{\bf y}_1 d{\bf y}_2
K_2({\bf x}_{p_2}-{\bf y}_1, {\bf x}_{p_2}-{\bf y}_2) 
\Big[ \widetilde{w}_2({\bf x}_{p_1},{\bf y}_{1}) \nonumber \\
& \times & \widetilde{w}_2({\bf x}_{p_3},{\bf y}_{2}) 
+\frac{1}{2} \widetilde{w}_4({\bf x}_{p_1},{\bf x}_{p_3}, 
{\bf y}_{1},{\bf y}_{2}) \Big]\, , \nonumber \\
\end{eqnarray}
where the sum is over all  permutations $p_1,p_2,p_3$ taking the values 
 $(1,2,3)$ and 
\begin{eqnarray}
\label{def1}
\widetilde{w}_2({\bf x}_1,{\bf x}_2) &\equiv&  
-\frac{1}{5} \langle (e^{-\zeta_1/5} -1) \zeta_2) 
\rangle_{\rm conn.}=-\frac{1}{5^2}e^{\langle \zeta^2\rangle/50}
\langle\zeta_1\zeta_2\rangle\, , \\
\label{def2}
\widetilde{w}_4({\bf x}_1,{\bf x}_2,{\bf x}_3,{\bf x}_4) &\equiv& 
\frac{1}{25} \langle(e^{-\zeta_1/5} -1) 
(e^{-\zeta_2/5} -1) \zeta_3 \zeta_4
\rangle_{\rm conn.} \\
&=&\frac{e^{\langle \zeta^2\rangle/25}}{3\times 5^4}
\left(
e^{\langle \zeta_1\zeta_2\rangle/25}-1\right)
\left(\langle\zeta_1\zeta_4\rangle
+\langle\zeta_2\zeta_4\rangle\right)\left(\langle\zeta_1\zeta_3\rangle
+\langle\zeta_2\zeta_3\rangle\right)+{\rm cyclic}\, .\nonumber
\end{eqnarray}
Despite the fact that we expanded the kernel up to
second-order, the expression ~(\ref{3point}) becomes exact at any order in
perturbation theory in the squezeed limit 
for which ${\cal K}$ tends to zero and for single field models
of inflation for which $a_{\rm NL}=1$. 
In such a case, the
exact three-point correlation function for the temperature anisotropies
on large-scales is
\begin{eqnarray}
W^{(3)}({\bf x}_1, {\bf x}_2, {\bf x}_3) &=& 
W_0^{(2)}({\bf x}_1,{\bf x}_2) 
W_0^{(2)}({\bf x}_1,{\bf x}_3) + W_0^{(2)}({\bf x}_1,{\bf x}_3) 
W_0^{(2)}({\bf x}_2, {\bf x}_3) \\
&+& W_0^{(2)}({\bf x}_1, {\bf x}_2) W_0^{(2)}({\bf x}_2, {\bf x}_3) 
+ W_0^{(2)}({\bf x}_1, {\bf x}_2) 
W_0^{(2)}({\bf x}_2, {\bf x}_3)
W_0^{(2)}({\bf x}_3, {\bf x}_1) \nonumber \, ,
\end{eqnarray}
where 
\begin{eqnarray}
W^{(2)}_0({\bf x}_i, {\bf x}_j) &\equiv& 
e^{\langle\zeta^2\rangle/50}
\left(e^{\langle\zeta_i\zeta_j\rangle/50}-1\right)\equiv
\int\frac{d^3k}{(2\pi)^3}\,e^{i{\bf k}\cdot({\bf x}_i-{\bf x}_j)}
P(k)\nonumber\\
&\simeq&
\frac{1}{50}
\int\frac{d\ln k}{2\pi^2}\,j_0\left(k\left|{\bf x}_i- {\bf x}_j\right|
\right)\,{\cal P}_\zeta(k)\, , 
\end{eqnarray}
and ${\cal P}_\zeta=A(k_0)^2(k/k_0)^{n_S-1}$ 
is the primordial power spectrum of the comoving
curvature perturbation with amplitude $A$ and spectral index $n_S$.
The expression for the exact bispectrum of temperature anisotropies 
valid at any order in perturbation theory is 
\begin{equation}
\Big< \frac{\delta_{np} T({\bf k}_1)}{T}
\frac{\delta_{np} T({\bf k}_2)}{T}\frac{\delta_{np} T({\bf k}_3)}{T}
\Big>=(2\pi)^3\,\delta^{(3)}\left({\bf k}_1+{\bf k}_2+{\bf k}_3\right)
\,B\left({\bf k}_1,{\bf k}_2, {\bf k}_3\right)\, ,
\end{equation}
where

\begin{eqnarray}
B\left({\bf k}_1,{\bf k}_2, {\bf k}_3\right)&=& 
P(k_1)P(k_2)+P(k_1)P(k_3)+P(k_2)P(k_3)\nonumber\\
&+&\int\frac{d {\bf q}}{(2\pi)^3}\,
P\left(\left|{\bf q}-{\bf k}_1\right|\right)
P\left(\left|{\bf q}-{\bf k}_2 \right|\right)
P\left(\left|{\bf q}-{\bf k}_3\right|\right)\\
&\simeq& 
2P(k_1)P(k_2)
+\int\frac{d {\bf q}}{(2\pi)^3}\,P\left(\left|{\bf q}\right|\right)
P\left(\left|{\bf q}-{\bf k}_2 \right|\right)
P\left(\left|{\bf q}+{\bf k}_2\right|\right)\, , (k_1\ll k_2,k_3)\nonumber \, .
\end{eqnarray}
For generic momenta configurations and for models for which
$a_{\rm NL}$ is sizeable,  
the exponentials present in the 
first line of Eq. (\ref{3point})  has to be  expanded
at second-order to consistently match the order of  the kernel $K$

\begin{equation}
\label{exp}
e^{-\zeta/5}\simeq 1-\frac{1}{5}\zeta +\frac{1}{2\times 5^2}\zeta^2+
{\cal O}\left(\zeta^3\right)\, .
\end{equation}
We immediately recover the expression obtained at second-order in
perturbation theory for the non-linearity parameter $f_{\rm NL}$ defined
as the coefficient of the gravitational potential $\Phi$ expanded
at second-order in terms of the linear Gaussian gravitational
potential $\Phi_{\rm L}=-\phi_1$, $\Phi=\Phi_{\rm L}+f_{\rm NL}\star
\left(\Phi_{\rm L}\right)^2$, with the convential Sachs-Wolfe effect
expressed as $\delta T/T=-(\Phi/3)$ \cite{prl}

\begin{equation}
\label{f}
f_{\rm NL} = -\left[\frac{5}{3}(1-a_{\rm NL})+\frac{1}{6}\right]
+ \left[3 ({\bf k}_1\cdot{\bf k}_3)
({\bf k}_2\cdot{\bf k}_3)/k^4-({\bf k}_1\cdot{\bf k}_2)/k^2\right]\, .
\end{equation}
It may be worth noticing that the coefficient 1/6 in the last expression
is simply the result of the expansion of the exponential (\ref{exp})
expressed in terms of the gravitational potential $\phi_1$: 
$e^{-\zeta/5}=e^{-\Phi_{\rm L}/3}\simeq 1-\frac{1}{3}\Phi_{\rm L}
+\frac{1}{18}\Phi^2_{\rm L}$, which gives the contribution $3\times(1/18)=1/6$
to $f_{\rm NL}$.

\subsection{The Trispectrum}
\noindent
Let us now follow a similar procedure to obtain the 4-point connected 
correlation function. In this case the kernel ${\mathcal K}$ appearing 
in Eq.~(\ref{final3}) must be expanded up to third order. In configuration 
space it can be written as a convolution
\begin{eqnarray}
\label{Kexpansion4p}
{\mathcal K}(\zeta)&=&\int d{\bf x}_1 d{\bf x}_2 K_2({\bf x}
-{\bf x}_1, {\bf x}-{\bf x}_2) \zeta({\bf x}_1)  
\zeta({\bf x}_2) \nonumber \\
&+&\int d{\bf x}_1 d{\bf x}_2 d{\bf x}_3 K_3({\bf x}
-{\bf x}_1, {\bf x}-{\bf x}_2, {\bf x}-{\bf x}_3) \zeta({\bf x}_1)  
\zeta({\bf x}_2) \zeta({\bf x}_3)\, ,
\end{eqnarray}  
where $K_2$ is the kernel defined by Eqs.~(\ref{Kexpansion3p}) and 
Eq.~(\ref{Kexpansion3p1}), while $K_3({\bf x}
-{\bf x}_1, {\bf x}-{\bf x}_2, {\bf x}-{\bf x}_3)$ is the triple inverse 
Fourier transform of the expression
\begin{eqnarray}
\label{K3}
\widetilde{K_3}({\bf k}_1,{\bf k}_2,{\bf k}_3)=(b_{\rm NL}-1)
+(a_{\rm NL}-1)\,{\mathcal A}({\bf k}_1,{\bf k}_2,{\bf k}_3)
+{\mathcal C}({\bf k}_1,{\bf k}_2,{\bf k}_3) \, ,
\end{eqnarray}
with
\begin{eqnarray}
\label{defA}
{\mathcal A}({\bf k}_1,{\bf k}_2,{\bf k}_3)&=& 
\frac{6}{5}\left[ \frac{{\bf k}_1 \cdot {\bf k}_4 ( 
{\bf k}_3 \cdot {\bf k}_4+{\bf k}_2 \cdot {\bf k}_4)+
({\bf k}_2 \cdot {\bf k}_4) ({\bf k}_3 \cdot {\bf k}_4)
}{k^4} \right. \nonumber \\
&-&\left. \frac{1}{3}
\frac{{\bf k}_1 \cdot ({\bf k}_2+{\bf k}_3)+{\bf k}_2 \cdot {\bf k}_3
}{k^2} \right]\, ,
\end{eqnarray}
\begin{eqnarray}
\label{defC}
{\mathcal C}({\bf k}_1,{\bf k}_2,{\bf k}_3) & = &
\frac{54}{25} \frac{
({\bf k}_4 \cdot {\bf k}_3)\,   
[({\bf k}_1+{\bf k}_2) \cdot {\bf k}_4]}{k^4}
\left[ \frac{
({\bf k}_1\cdot({\bf k}_1+{\bf k}_2))\, 
({\bf k}_2\cdot({\bf k}_1+{\bf k}_2))}{
|{\bf k}_1+{\bf k}_2|^4} \right. \nonumber \\
&-& \left. \frac{1}{3} \frac{{\bf k}_1\cdot{\bf k}_2}
{|{\bf k}_1+{\bf k}_2|^2} \right] +{\rm cycl.}\, ,
\end{eqnarray}
where $k=|{\bf k}_4|$ and ${\bf k}_4=-({\bf k}_1+{\bf k}_2+{\bf k}_3)$. In 
Eq.~(\ref{defC}) one has to take cyclic terms by an exchange of the 
wavenumbers ${\bf k}_1, {\bf k}_2, {\bf k}_3$. 

In order to compute 
$\widetilde{K_3}({\bf k}_1,{\bf k}_2,{\bf k}_3)$ we have applied the 
iterative procedure described in Section IV taking into account also a 
possible primordial non-Gaussian contribution by expanding the curvature 
perturbation as

\begin{equation}
\zeta=\zeta_{\rm L}+ (a_{\rm NL}-1) 
\star \zeta_{\rm L}^2+(b_{\rm NL}-1)\star \zeta_{\rm L}^3\, . 
\end{equation}
The value of $b_{\rm NL}$ will depend 
on the different scenarios for the generation of the cosmological 
perturbations. For example, for standard single-field models of inflation 
$b_{\rm NL} = 1$ (plus tiny contributions proportional to powers of the 
slow-roll parameters), while for other scenarios it might well be 
non-negligible. For simplicity from now on we will remove the subscript ``L''.

Similarly to the bispectrum, also for the 4-point connected correlation 
function there exists a specific limit for which the kernel ${\mathcal K}$ 
tends to zero, corresponding to take two wavenumbers much smaller than the 
other ones, {\it e.g.} $k_1,k_2 \ll k_3, k_4$. For this limit and in the 
case of single-field models of inflation ($a_{\rm NL}=1, b_{\rm NL}=1$), one 
can compute an exact expression of the trispectrum by Fourier 
transforming the 4-point connected correlation function 
which, in this limit, reads
\begin{eqnarray}
& & W^{(4)}({\bf x}_1, {\bf x}_2, {\bf x}_3, {\bf x}_4)
= \langle (e^{-\zeta({\bf x}_1)/5}-1) 
(e^{-\zeta({\bf x}_2)/5}-1) (e^{-\zeta({\bf x}_3)/5}-1) 
(e^{-\zeta({\bf x}_4)/5}-1)  
\rangle_{\rm conn.}\nonumber \\
& & =
W_0^{(2)}({\bf x}_1,{\bf x}_2) W_0^{(2)}({\bf x}_1,{\bf x}_3) 
W_0^{(2)}({\bf x}_1,{\bf x}_4) W_0^{(2)}({\bf x}_2,{\bf x}_3) 
W_0^{(2)}({\bf x}_2,{\bf x}_4) 
W_0^{(2)}({\bf x}_3,{\bf x}_4) \nonumber \\
& &+W_0^{(2)}({\bf x}_1,{\bf x}_4)W_0^{(2)}({\bf x}_2,{\bf x}_3)
W_0^{(2)}({\bf x}_2,{\bf x}_4)W_0^{(2)}({\bf x}_3,{\bf x}_4)\left(
W_0^{(2)}({\bf x}_1,{\bf x}_2)+W_0^{(2)}({\bf x}_1,{\bf x}_3) \right)
\nonumber \\
& & +W_0^{(2)}({\bf x}_1,{\bf x}_2)W_0^{(2)}({\bf x}_1,{\bf x}_3)
W_0^{(2)}({\bf x}_2,{\bf x}_4)W_0^{(2)}({\bf x}_3,{\bf x}_4)\left(
W_0^{(2)}({\bf x}_1,{\bf x}_4)+W_0^{(2)}({\bf x}_2,{\bf x}_3) \right)
\nonumber \\
& & +W_0^{(2)}({\bf x}_2,{\bf x}_3)W_0^{(2)}({\bf x}_2,{\bf x}_4)
W_0^{(2)}({\bf x}_3,{\bf x}_4)\left(
 W_0^{(2)}({\bf x}_1,{\bf x}_2)+W_0^{(2)}({\bf x}_1,{\bf x}_3)+
W_0^{(2)}({\bf x}_1,{\bf x}_4) \right)\nonumber \\
& & +W_0^{(2)}({\bf x}_1,{\bf x}_4)W_0^{(2)}({\bf x}_2,{\bf x}_4)
W_0^{(2)}({\bf x}_3,{\bf x}_4) \left(W_0^{(2)}({\bf x}_1,{\bf x}_2)+
W_0^{(2)}({\bf x}_1,{\bf x}_3)\right)\nonumber \\
& & +
\left(W_0^{(2)}({\bf x}_2,{\bf x}_4)+ W_0^{(2)}({\bf x}_3,{\bf x}_4)\right) \left( W_0^{(2)}({\bf x}_1,{\bf x}_2) W_0^{(2)}({\bf x}_1,{\bf x}_4) 
W_0^{(2)}({\bf x}_2,{\bf x}_3)+ \right. \nonumber \\
& & \left. + 
W_0^{(2)}({\bf x}_2,{\bf x}_3) W_0^{(2)}({\bf x}_1,{\bf x}_4) 
W_0^{(2)}({\bf x}_1,{\bf x}_3) + 
W_0^{(2)}({\bf x}_1,{\bf x}_3) W_0^{(2)}({\bf x}_2,{\bf x}_3) 
W_0^{(2)}({\bf x}_1,{\bf x}_2) \right) \nonumber \\
& & +W_0^{(2)}({\bf x}_1,{\bf x}_3) W_0^{(2)}({\bf x}_1,{\bf x}_4) 
W_0^{(2)}({\bf x}_1,{\bf x}_2) \left(W_0^{(2)}({\bf x}_2,{\bf x}_3)+
W_0^{(2)}({\bf x}_2,{\bf x}_4)+W_0^{(2)}({\bf x}_3,{\bf x}_4) \right) \nonumber \\
& & + W_0^{(2)}({\bf x}_1,{\bf x}_2) W_0^{(2)}({\bf x}_1,{\bf x}_3) 
W_0^{(2)}({\bf x}_2,{\bf x}_4) W_0^{(2)}({\bf x}_3,{\bf x}_4) \nonumber \\
& & +W_0^{(2)}({\bf x}_1,{\bf x}_2) W_0^{(2)}({\bf x}_1,{\bf x}_3) \left(
W_0^{(2)}({\bf x}_1,{\bf x}_4)+W_0^{(2)}({\bf x}_2,{\bf x}_4)+
W_0^{(2)}({\bf x}_3,{\bf x}_4)
\right) \nonumber \\
& & +
W_0^{(2)}({\bf x}_1,{\bf x}_4)W_0^{(2)}({\bf x}_2,{\bf x}_3) \left(
W_0^{(2)}({\bf x}_1,{\bf x}_3)+W_0^{(2)}({\bf x}_2,{\bf x}_4)+
W_0^{(2)}({\bf x}_3,{\bf x}_4)
\right) \nonumber \\
& & + 
W_0^{(2)}({\bf x}_2,{\bf x}_4)W_0^{(2)}({\bf x}_3,{\bf x}_4) \left(
W_0^{(2)}({\bf x}_1,{\bf x}_2)+W_0^{(2)}({\bf x}_1,{\bf x}_3)+
W_0^{(2)}({\bf x}_1,{\bf x}_4) \right)\nonumber \\
& & +
\left(W_0^{(2)}({\bf x}_1,{\bf x}_2)+ W_0^{(2)}({\bf x}_1,{\bf x}_3)
\right) \left( 
W_0^{(2)}({\bf x}_2,{\bf x}_3)W_0^{(2)}({\bf x}_2,{\bf x}_4)+
W_0^{(2)}({\bf x}_2,{\bf x}_3)W_0^{(2)}({\bf x}_3,{\bf x}_4)
\right) \nonumber \\
& &+
W_0^{(2)}({\bf x}_1,{\bf x}_2)W_0^{(2)}({\bf x}_1,{\bf x}_4)
W_0^{(2)}({\bf x}_3,{\bf x}_4)+
W_0^{(2)}({\bf x}_1,{\bf x}_3)W_0^{(2)}({\bf x}_1,{\bf x}_4)
W_0^{(2)}({\bf x}_2,{\bf x}_4) \, .
\end{eqnarray}
On the other hand, 
for generic momenta configurations and for models for which $a_{\rm NL}$ 
and $b_{\rm NL}$ are sizeable we can expand the exponential entering in the 
expression~(\ref{final3}) of the temperature anisotropies up to third order 
and use the kernel ${\mathcal K}(\zeta)$ in Eq.~(\ref{Kexpansion4p}). 
In this way we are able to determine the 
non-linearity parameter $g_{\rm NL}$ which enters into the trispectrum of 
the CMB anisotropies according, for example, to the analysis of 
Refs.~\cite{hu,jul}. 
The parameter $g_{\rm NL}$ is defined through the expansion 
of the (Bardeen) gravitational potential $\Phi$ up to third-order as 
\begin{equation}
\Phi=\Phi_L+f_{\rm NL} \star (\Phi_L)^2+ g_{\rm NL} \star (\Phi_L)^3\, ,
\end{equation}
where $\Phi_L=-\phi_1$, is the linear Gaussian part of $\Phi$. 
We find the following expression 
\begin{eqnarray}
g_{\rm NL}({\bf k_1},{\bf k}_2,{\bf k_3})& = &\frac{25}{9} (b_{\rm NL}-1)+
\frac{25}{9}(a_{\rm NL}-1) {\mathcal A}({\bf k}_1,{\bf k}_2,{\bf k}_3)+
\frac{25}{9}{\mathcal C}({\bf k}_1,{\bf k}_2,{\bf k}_3)
-\frac{5}{9}(a_{\rm NL}-1) \nonumber \\
&+&\frac{1}{54} 
-\frac{1}{3} 
\left[ 
\frac{
({\bf k}_1\cdot({\bf k}_1+{\bf k}_2))\,
({\bf k}_2\cdot({\bf k}_1+{\bf k}_2))}{
|{\bf k}_1+{\bf k}_2|^4}-
\frac{1}{3} \frac{{\bf k}_1\cdot{\bf k}_2}
{|{\bf k}_1+{\bf k}_2|^2}+ {\rm cycl.}\right]\, ,
\end{eqnarray}      
where $\mathcal{A}({\bf k}_1,{\bf k}_2,{\bf k}_3)$ and 
$\mathcal{C}({\bf k}_1,{\bf k}_2,{\bf k}_3)$
are defined through Eqs.~(\ref{defA}) 
and~(\ref{defC}).

\section{Conclusions}
\noindent
In this paper we showed how to calculate exactly 
the $n$-point correlation function
of CMB anisotropies in the case in which all wavelengths are beyond the
horizon at last scattering. In this limit 
the Sachs-Wolfe effect is 
predominant and its contribution  to higher-order correlation functions
yields the most direct signal of non-Gaussianity in the primordial
cosmological seeds. Of particular interest are the bispectrum and the
trispectrum which can be used to assess the level of 
non-Gaussianity on cosmological scales. We have calculated 
the non-perturbative expressions for the
bispectrum and the trispecturm as predicted within 
single-field models of inflation and in the so-called ``squeezed'' limit
in which some of the wavenumbers are much smaller than the others. 
For other scenarios of generation of the cosmological perturbations, we 
have  provided the non-linearity parameters $f_{\rm NL}$ and
$g_{\rm NL}$ entering respectively the theoretical predictions
of the bispectrum and trispectrum. Our results for the bispectrum and the 
trispectrum represent the essential input in order to obtain the predicted 
angular bispectrum $B_{l_1 l_2 l_3}=\langle a_{l_1 m_1} a_{l_2 m_2} a_{l_3 m_3} 
\rangle$ and the trispectrum of CMB anisotropies 
(and higher order correlation functions)
according, for example, to the formalism developed in Refs.~\cite{SG,wk}. In 
these works it is shown how to compute the angular 
bispectrum accounting for a non-trivial 
wavenumber dependence of the non-linearity parameter 
$f_{\rm NL}({\bf k}_1,{\bf k}_2)$. 
The angular modulation of the quadratic non-linearity 
predicted by Eq.~(\ref{f}) 
is currently under investigation~\cite{LMR}, adopting 
the technique of Ref.~\cite{SG}, in order to look for specific signatures of 
inflationary non-Gaussianity in the CMB. 
Notice that for a $\Lambda$CDM cosmology a 
late integrated Sachs-Wolfe effect arising from 
the explicit  time dependence of the linear gravitational potential
during the late accelerated phase would also give a contribution on large scales. 
The formalism developed in this paper can be extended to take into account also 
for this effect. On the other hand on smaller scales there will be other  
effects contributing to CMB non-Gaussianity such as gravitational lensing, 
Shapiro time-delays and Rees-Sciama effects 
produced at the non-linear level. One should be able 
to compute the angular connected correlation functions induced by these effects  
by using the techinique developed in Ref.~\cite{CZ} and to 
distinguish them from the large-scale Sachs-Wolfe effect provided here thanks to 
their specific angular dependence. Our predictions for the
higher-order correlation functions should be compared model
by model with the unavoidable contributions 
from various secondary anisotropies 
and systematic effects, such as astrophysical foregrounds.

\vspace{1cm}
\section{Appendix}
\setcounter{equation}{0}
\def\theequation{A.\arabic{equation}}
\vskip 0.2cm
\label{A}
\noindent 
In this Appendix we will show in detail how to compute the $n$-point
 connected 
correlation functions~(\ref{def2}) by making use of the generating functional 
$w(\bf{x}_1,...,\bf{x}_n)$ and in particular how to get the 
result~~(\ref{3point}). 

Let us consider the  quantity
\begin{equation}
\delta_T=e^{\varphi+K(\varphi)}-1\, ,
\end{equation}
where $\varphi$ is a generic Gaussian random field, and $K(\varphi)$ is a 
generic 
functional of $\varphi$ (apart from multiplicative coefficients $\varphi$ will
be identified with  the comoving curvature perturbation  
$\zeta$ and $K$ will be given by the kernel ${\mathcal K}$ through 
the iterative procedure). 

The generating functional for the correlated functions of $\delta_T$
 is given by
\begin{equation}
Z[J]=\int{\mathcal D}[\varphi] {\mathcal P}[\varphi] e^{i\int d{\bf x} 
J({\bf x}) (e^{\varphi+K(\varphi)}-1)}\, ,
\label{ooo}
\end{equation}
where $J({\bf x})$ is an arbitrary external 
source. The functional integral is over 
all the $\varphi$ configurations weighted by the Gaussian probability density 
functional 
\begin{equation}
\label{defPG}
{\mathcal P}[\varphi]=\frac{e^{-\frac{1}{2}\int d{\bf y} d{\bf x} 
\varphi({\bf y}) {\mathcal G}({\bf y},{\bf x})\varphi({\bf x})}}{
\int {\mathcal D}[\varphi] e^{-\frac{1}{2}\int d{\bf y} d{\bf x} 
\varphi({\bf y}) {\mathcal G}({\bf y},{\bf x})\varphi({\bf x})}  }\, ,
\end{equation}
which has been properly normalized in such a way that the total probability 
equals unity, $\int {\mathcal D}[\varphi] {\mathcal P}[\varphi] =1$. 

To compute the functional 
derivatives with respect to $J$ we find it convenient to use an additional 
arbitrary source $\lambda({\bf x})$.  We 
introduce the following generating 
functional
\begin{equation}
\label{WJL}
Z[J,\lambda]
=\int{\mathcal D}[\varphi] {\mathcal P}[\varphi] e^{i\int d{\bf x} 
J({\bf x}) (e^{\varphi+K(\varphi)}-1)} 
e^{i\int d{\bf x} \lambda({\bf x}) \varphi({\bf x})} \, ,
\end{equation}
which reduces to the expression (\ref{ooo}) when $\lambda=0$. 
Functionals of the form  in Eq.~(\ref{WJL}) are common  in 
field theory when computing correlation functions of 
{\it composite operators} 
(in which 
 case $J({\bf x})$ represents a ``local coupling''; see for example 
Ref.~\cite{Zinn}).
The correlation functions generated by $Z[J,\lambda]$ are given by 
\begin{eqnarray}
\langle  (e^{\varphi({\bf y}_1)+K(\varphi({\bf y}_1))}-1)...
(e^{\varphi({\bf y}_n)+
K(\varphi({\bf y}_n))}-1)\varphi({\bf x}_1)...\varphi({\bf x}_m) 
\rangle  \nonumber \\
= i^{-n+m}
\frac{\delta^{n+m} Z[J,\lambda]}{\delta J({\bf y}_1)...
\delta J({\bf y}_n)
\delta \lambda({\bf x}_1)....\delta \lambda({\bf x}_m)} 
\Bigg|_{(J,\lambda)=0} \, . \nonumber \\
\end{eqnarray}
Thus we will take only derivatives with respect to $J$ to get the correlation 
functions of $e^{\varphi+K(\varphi)}$.
If we now write 
\begin{equation}
\label{exp0}
e^{\varphi+K(\varphi)}-1=\sum_{n=1}^{\infty}\frac{K^n(\varphi)}{n!}e^{\varphi}+
(e^{\varphi}-1)\, ,
\end{equation}
we can rewrite Eq.~(\ref{WJL}) as 
\begin{equation}
\label{exp1}
Z[J,\lambda]=e^{i\int d{\bf x} J({\bf x}) 
\left( \sum_{n=1} \frac{K^n(\frac{1}{i}\frac{\delta}{\delta \lambda})}
{n!} e^{\frac{1}{i}\frac{\delta}{\delta \lambda}} \right)} Z_0[J,\lambda]\, ,
\end{equation} 
where $Z_0[J,\lambda]$ corresponds to the generating functional 
$Z[J,\lambda]$ when the function $K(\varphi)=0$
\begin{eqnarray}
Z_0[J,\lambda] & = &
\int{\mathcal D}[\varphi] {\mathcal P}[\varphi] e^{i\int d{\bf x} 
J({\bf x}) (e^{\varphi}-1)} 
e^{i\int d{\bf x} \lambda({\bf x}) \varphi({\bf x})} \nonumber \\
&\equiv&  e^{W_0[J,\lambda]}\, .
\end{eqnarray}
In writing Eq.~(\ref{exp1}) we have made use of the  property 
$i^{-1} \delta e^{i \int d{\bf y} \lambda({\bf y}) 
\varphi({\bf x})}/\delta J({\bf y}) = \varphi({\bf x}) e^{i \int d{\bf y} 
\lambda({\bf y}) \varphi({\bf y})}$, 
in order to isolate the 
``interaction term'' (see, for example, Ref.~\cite{Ramond}).

Now let us write
\begin{equation}
\label{finit}
W[J,\lambda]\equiv \ln Z[J,\lambda]=W_0[J,\lambda]
+ \ln \Big[1 
+e^{-W_0} \left( e^{i \int d{\bf x} J({\bf x}) 
\left( \sum_{n=1} \frac{K^n(\frac{1}{i}\frac{\delta}{\delta \lambda})}
{n!} e^{\frac{1}{i}\frac{\delta}{\delta \lambda}} \right)}-1\right)
e^{W_0} \Big] \, .\nonumber \\ 
\end{equation}
The derivatives with respect to $J$ evaluated for $(J,\lambda)=0$ will give 
the connected correlation functions we are looking for, accounting also for 
the kernel $K$. We are using the standard procedure to evaluate the 
connected correlation functions, for example, for an interacting scalar field 
(see e.g. Ref.~\cite{Ramond}), except that 
in our case the ``interaction term'' is related to the kernel $K$ and 
we will make a perturbative expansion around 
$W_0[J,\lambda]$, since 
the derivatives of $W_0[J,\lambda]$ 
with respect to $J$ (evaluated for $(J,\lambda)=0$)   
give the connected correlated functions 
for $\left(e^{\varphi}-1\right)$, see Eq.~(\ref{path}).
  
At this point we have to perform a perturbative expansion. We suppose 
that the perturbation parameter is the small {\rm r.m.s} 
amplitude of the perturbations, 
$\varphi_{\rm rms} \ll 1$ and  use the fact that the function 
$K(\varphi)$ is obtained through the iterative procedure described in  
Section III. Let us suppose  that 
\begin{equation}
\label{Kiter}
K(\varphi)=a * \varphi^2+b * \varphi^3+.... \, ,
\end{equation}
where the star denotes  a 
convolution operation in configuration space. 
For the sake of simplicity, in the following 
we will neglect the convolutions, and treat $a,b,..$ 
as constant coefficients. 
In Eq.~(\ref{finit}) we will focus on the term $\ln(1+\Delta)$ where the 
definition of $\Delta$ is obtained by comparison with Eq.~(\ref{finit}). Thus 
$\ln (1+\Delta) =1+\Delta-\Delta^2/2+\Delta^3/3+...$.\\
Now let us consider how to compute, for example, the three-point correlated 
function given in Eq.~(\ref{3point}). At lowest order in our approximation, 
the only term that in this case is relevant is just $\Delta$ and moreover 
we keep only the first term coming from the expansion of the exponential
\begin{equation}
\Delta \simeq  
e^{-W_0} i \int d{\bf x} J({\bf x}) 
\left( \sum_{n=1} \frac{K^n(\frac{1}{i}\frac{\delta}{\delta \lambda})}
{n!} e^{\frac{1}{i}\frac{\delta}{\delta \lambda}} \right)
e^{W_0} \, .
\end{equation}
Here we have to expand once more, and looking at Eq.~(\ref{Kiter}) 
we just need for the bispectrum 
to take $K=a \varphi^2$, and from  
$(e^{\frac{1}{i}\frac{\delta}{\delta \lambda}} =1 +\delta/(i \delta \lambda)
+....)$ we just pick up the factor 1..\\
Thus we are left with 
\begin{eqnarray}
\label{left}
\ln (1+\Delta) &\simeq& 
\Delta \simeq i a e^{-W_0} \int d{\bf x} J({\bf x}) 
\left( \frac{1}{i} \frac{\delta}{\delta \lambda} \right)^2 e^{W_0}
\nonumber \\
&=& i^{-1} a \int d{\bf x} J({\bf x}) \left( 
\frac{\delta W_0}{\delta \lambda ({\bf x})} \right)^2
+\frac{\delta^2 W_0}{\delta \lambda^2({\bf x})} \, . \nonumber \\
\end{eqnarray}
Therefore the bispectrum is given by 
\begin{eqnarray}
\label{Z3}
& &W^{(3)}({\bf y}_1,{\bf y}_2, {\bf y}_3) = 
i^{-3} \frac{\delta^3 W[J,\lambda]}{\delta J({\bf y}_1) 
\delta J({\bf y}_2) \delta J({\bf y}_3)} \Bigg|_{(J,\lambda)=0}\nonumber \\
& &= \langle (e^{\varphi({\bf y}_1)}-1) (e^{\varphi({\bf y}_2)}-1)                   (e^{\varphi({\bf y}_3)}-1)    \rangle_{\rm conn.} \nonumber \\
& & +i^{-3} \frac{\delta^3 \Delta[J,\lambda]}
{\delta J({\bf y}_1) \delta J({\bf y}_2) \delta J({\bf y}_3)} 
\Bigg|_{(J,\lambda)=0}\, ,
\end{eqnarray}
where we have to compute the last contribution in Eq.~(\ref{Z3}).
We need an expression for 
$W_0[J,\lambda]$. Since the derivatives 
of $W_0$ with respect to $J$ and $\lambda$ 
(evaluated in $(J,\lambda)=0$) give the connected 
correlation functions for $(e^{\varphi}-1)$ and $\varphi$, we can write
\begin{equation}
W_0[J,\lambda]=\sum_{n=1}^{\infty}\frac{i^n}{n!} \int 
d{\bf x}_1....d{\bf x}_n \widetilde{z}({\bf x}_1....{\bf x}_n) 
\widetilde{J}({\bf x}_1)...\widetilde{J}({\bf x}_n) \, ,
\end{equation}
where $\widetilde{J}({\bf x}_i)$ can be either the source ${J}({\bf x}_i)$ or  
$\lambda({\bf x}_i)$, and correspondingly 
$\widetilde{w}_n({\bf x}_1....{\bf x}_n)$ are the connected correlation 
functions. For example $\langle (e^{\varphi({\bf x}_1)}-1) 
\varphi({\bf x}_2) \varphi({\bf x}_3) \rangle$ for the choice ${J}({\bf x}_1)
\lambda({\bf x}_2) \lambda({\bf x}_3)$, and one has to consider
different combinations. 

Using the usual operations for functional derivatives 
(see, for example,~\cite{Mosel}) we find  
\begin{equation}
\label{result1}
i^{-3} \frac{\delta^3 \Delta[J,\lambda]}
{\delta J({\bf y}_1) \delta J({\bf y}_2) \delta J({\bf y}_3)} 
\Bigg|_{(J,\lambda)=0}= 
 a  \sum_p \Big[\widetilde{w}_2({\bf y}_{p_1},
{\bf y}_{p_2}
\widetilde{w}_2({\bf y}_{p_3},{\bf y}_{p_2}) 
+\frac{1}{2} \widetilde{z}_4({\bf y}_{p_1},{\bf y}_{p_2}, 
{\bf y}_{p_3},{\bf y}_{p_3}) \Big] 
\end{equation}
where the sum is over the permutations $p_1,p_2,p_3$ of  indices 
$(1,2,3)$ and 
we have used the following notations
\begin{eqnarray}
\label{Adef1}
\widetilde{w}_2({\bf x},{\bf y}) &\equiv& \langle 
(e^{\varphi({\bf x})} -1) \varphi({\bf y}) \rangle_{\rm conn.} \\
\label{Adef2}
\widetilde{w}_4({\bf x}_1,{\bf x}_2,{\bf x},{\bf x}) &\equiv& 
\langle (e^{\varphi({\bf x}_1)} -1) 
(e^{\varphi({\bf x}_2)} -1) \varphi({\bf x}) \varphi({\bf x})
\rangle_{\rm conn}\, . \nonumber 
\end{eqnarray}
Notice that it is possible to compute the connected correlation functions 
appearing in Eq.~(\ref{Adef1}) and~(\ref{Adef2}) in a similar manner to what 
one does for $e^{\varphi}$.

The result in Eq.~(\ref{result1}) is a sum of two terms that 
correspond to the two pieces in Eq.~(\ref{left}). One has two 
evaluate $\varphi$ at two equal points because 
we have taken the derivative twice with respect to 
$\lambda$. Finally the product of the two-point 
connected correlation functions and the presence of the 
fourth-order  connected correlation 
function are due to the fact that we have the product of two first 
derivatives {\it w.r.t} $\lambda$ and a second order derivative 
{\it w.r.t} 
$\lambda$, respectively. 
Taking then the three derivatives {\it w.r.t.} $J$ involves 
$(e^{\varphi}-1)$ in the correlations.
Now it is easy to generalize the result~(\ref{result1}) when $a$ is not a 
constant coefficient but we have a kernel in configuration space such that
\begin{equation}
K(\varphi)=\int d{\bf {\bar x}}_1 d{\bf {\bar x}}_2 K({\bf x}
-{\bf {\bar x}_1},
{\bf x}-{\bf {\bar x}_2}) \varphi({\bf {\bar x}}_1)  
\varphi({\bf {\bar x}}_2)+....
\end{equation} 
As one can guess from Eq.~(\ref{result1}) one has 
\begin{eqnarray}
\label{result2}
& i^{-3}& \frac{\delta^3 \Delta[J,\lambda]}
{\delta J({\bf y}_1) \delta J({\bf y}_2) \delta J({\bf y}_3)} 
\Bigg|_{(J,\lambda)=0} = \nonumber \\ 
& \sum_p &  
\int d{\bf {\bar x}}_1 d{\bf {\bar x}}_2 
K({\bf y}_{p_2}-{\bf {\bar x}_1}, {\bf y}_{p_2}-{\bf {\bar x}_2}) 
\widetilde{w}_2({\bf y}_{p_1},{\bf {\bar x}}_{1})
\widetilde{w}_2({\bf y}_{p_3},{\bf {\bar x}}_{2}) \nonumber \\
&+&   \frac{1}{2} \sum_p
\int d{\bf {\bar x}}_1 d{\bf {\bar x}}_2 
K({\bf y}_{p_3}-{\bf {\bar x}_1}, {\bf y}_{p_3}-{\bf {\bar x}_2}) 
\widetilde{w}_4({\bf y}_{p_1},{\bf y}_{p_2}, 
{\bf {\bar x}}_{1},{\bf {\bar x}}_{2}) \, . \nonumber \\
\end{eqnarray}
In fact we have explicitly verified that this is the correct generalization 
of Eq.~(\ref{result1}).\\
Thus from Eq.~(\ref{finit}) and Eq.~(\ref{result2}) 
the connected three-point correlation function for 
$(e^{\varphi+K(\varphi)}-1)$ reads  
\begin{eqnarray}
\label{result3}
& & W^{(3)}({\bf y}_1,{\bf y}_2,{\bf y}_3)=
i^{-3} \frac{\delta^3 Z[J,\lambda]}{\delta J({\bf y}_1) 
\delta J({\bf y}_2) \delta J({\bf y}_3)} \Bigg|_{(J,\lambda)=0}= \nonumber \\ 
& & \langle (e^{\varphi({\bf y}_1)}-1) 
(e^{\varphi({\bf y}_2)}-1) (e^{\varphi({\bf y}_3)}-1) \rangle_{\rm connected}
\nonumber \\
&+& \sum_p  
\int d{\bf {\bar x}}_1 d{\bf {\bar x}}_2 
K({\bf y}_{p_2}-{\bf {\bar x}_1}, {\bf y}_{p_2}-{\bf {\bar x}_2}) 
\widetilde{w}_2({\bf y}_{p_1},{\bf {\bar x}}_{1})
\widetilde{w}_2({\bf y}_{p_3},{\bf {\bar x}}_{2}) \nonumber \\
&+&\frac{1}{2} \sum_p
\int d{\bf {\bar x}}_1 d{\bf {\bar x}}_2 
K({\bf y}_{p_3}-{\bf {\bar x}_1}, {\bf y}_{p_3}-{\bf {\bar x}_2}) 
\widetilde{w}_4({\bf y}_{p_1},{\bf y}_{p_2}, 
{\bf {\bar x}}_{1},{\bf {\bar x}}_{2}) \, , \nonumber \\
\end{eqnarray}
where one has to use the definitions in Eqs.~(\ref{Adef1}) and~(\ref{Adef2}).

\section*{Acknowledgments}
N.B. would like to thank James Babington for useful discussions on 
techniques of functional-integral analysis in quantum field theory.



\end{document}